\documentclass[fleqn,twoside]{article}
\usepackage{espcrc2}
\usepackage{graphicx}

\newcommand{\AmS}{{\protect\the\textfont2
  A\kern-.1667em\lower.5ex\hbox{M}\kern-.125emS}}

\title{Where do perturbative and non-perturbative QCD meet?}

\author{G.S.\ Bali\address[GLA]{Department of Physics and Astronomy,
University
of Glasgow, Glasgow G12 8QQ, UK},
P.\ Boyle\address{Department of Physics and Astronomy,
University of Edinburgh, Edinburgh
EH9 3JZ, UK}\thanks{Present address: Physics Department, Columbia University,
New York, NY 10027, USA}
and C.T.H.\ Davies\addressmark[GLA]
}
       
\begin{document}

\begin{abstract}
We computed the static potential and Wilson loops
to $O(\alpha^2)$ in perturbation theory for different lattice
quark and gluon actions. In general, we find short distance lattice
data to be well described
by ``boosted perturbation theory''.
For Wilson-type fermions at present-day quark masses and
lattice spacings agreement within 10~\%
between measured ``$\beta$-shifts'' and those predicted
by perturbation theory is found.
We comment on prospects for a determination of the real world
QCD running coupling.
\end{abstract}

\maketitle

\section{INTRODUCTION}
Perturbative results on
{\em physical} quantities
are useful in lattice simulations in several ways:\\
$\bullet$ to qualitatively
understand ``improvement''
and ``lattice artefacts'',\\
$\bullet$ to numerically
validate that the continuum limit exists and is
reached as $\beta\rightarrow\infty$,\\
$\bullet$ for predicting the ``$\beta$-shift'',
resulting from massive sea quarks,\\
$\bullet$ to estimate quantities that are hard to obtain otherwise, e.g.\
the static quark self energy,\\
$\bullet$ to calculate $\alpha_s$ from low energy QCD phenomenology.

In view of this we calculated Wilson loops and the static potential
with massive Wilson, Sheikholeslami-Wohlert (SW) and
Kogut-Susskind (KS) fermions to $O(\alpha^2)$~\cite{inprep}.
We define the potential,
$
aV_{\mbox{\scriptsize int}}({\mathbf R}a)=v_{L,1}({\mathbf R})\alpha_L
+v_{L,2}({\mathbf R})\alpha_L^2+\cdots,
$
where $V_{\mbox{\scriptsize int}}({\mathbf R}a)=V({\mathbf R}a)-V(\infty)$.
The static quark self energy,
$V(\infty)=2\,\delta m_{\mbox{\scriptsize stat}}\propto \alpha a^{-1}+\cdots$,
vanishes in dimensional regularisation but diverges on the lattice as
the continuum limit is approached. For $SU(3)$ we find,
$a\delta m_{\mbox{\scriptsize stat}}=2.1172743\,
\alpha_L+[11.143(3)+n_fY_f]\alpha_L^2$
with $Y_f=-0.36846(6)$ for massless KS quarks and
$Y_f=-0.42333(6)+0.0516(2)c_{SW}-0.5870(2)c_{SW}^2$ for Wilson-SW
quarks. While our value of the Wilson-SW
$Y_f$ agrees with Ref.~\cite{Martinelli:1999vt}
we find a $4\,\sigma$ discrepancy in
the gluonic contribution.
\section{``$\Delta K_1$'' AND THE ``$\beta$-SHIFT''}
\begin{figure}
\includegraphics[width=17pc]{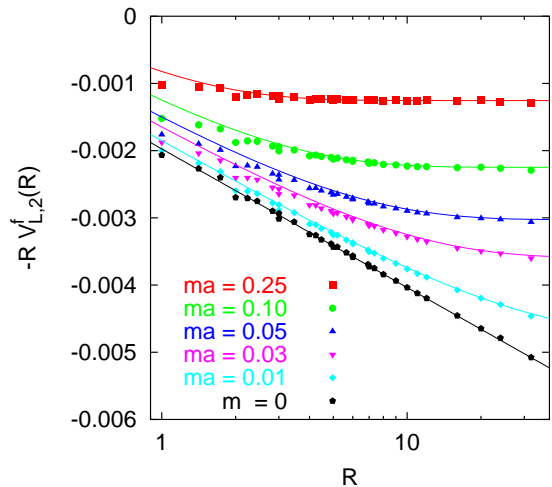}
\caption{Wilson quark contribution to $v_{L,2}$.\vskip -.7cm}
\label{fig:massive}
\end{figure}
The $\overline{MS}$ scheme is related to the lattice scheme via
$\alpha_{\overline{MS}}(a^{-1})=\alpha_L+b_1\alpha^2+\cdots$,
with the conversion factor
$b_1=-\pi/(2N)+k_1N+K_1(ma)n_f$,
where the numerical constant $k_1$ is known for a variety
of gluonic actions and $K_1(0)$ is known for Wilson, SW and KS quarks with
Wilson glue. Further couplings can be defined, e.g.\ from the
potential in position space, $\alpha_R(\mu)=
-rV_{\mbox{\scriptsize int}}(r)/C_F$, $\mu=r^{-1}$.
While the $\overline{MS}$ scheme is ``mass-independent'',
the $\beta$-function coefficients $\beta_i$ of the above
``$R$'' scheme
depend on $m/\mu$ and hence on the quark mass; universality is lost
and the $\beta_i$ are specific for each dimensionful observable
that is studied.
The same holds true for the
conversion factor to the $\overline{MS}$ scheme.

\begin{figure}
\includegraphics[width=17pc]{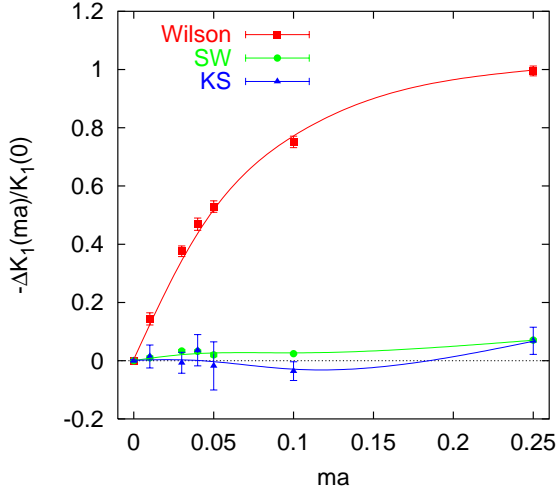}
\caption{Matching with the $\overline{MS}$ scheme.\vskip -.7cm}
\label{fig:shift}
\end{figure}
\begin{figure}
\includegraphics[width=17pc]{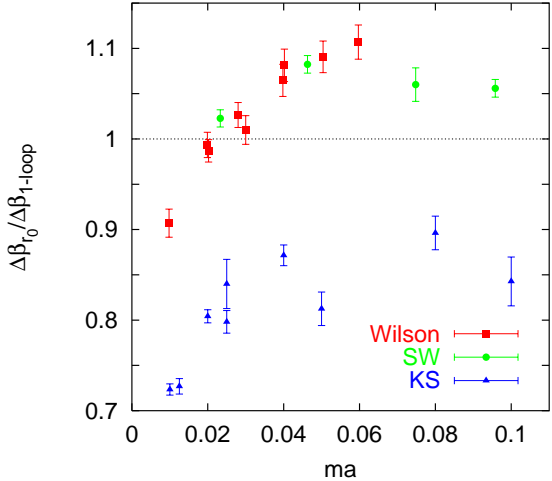}
\caption{Comparison with the measured $\beta$-shift.\vskip -.7cm}
\label{fig:compbshift}
\end{figure}

In contrast,
the lattice scheme is mass-independent in the continuum limit.
However, it is often worthwhile to study quantities that are not
defined in this limit like small Wilson loops~\cite{Davies:1997mg},
and some effective field theory
approaches require a finite lattice cut-off too.
At fixed physical
$r=Ra$ the limit $R\rightarrow\infty$ corresponds to the continuum
limit $a\rightarrow 0$, in which rotational symmetry is restored and
lattice and continuum perturbative predictions agree.
One would also expect rotational symmetry to be
restored at finite $a$ for distances $r\gg a$. These two cases
become distinguishable in perturbation
theory once an external scale $m$ is introduced and hence the
situation $r< m$ differs from $r> m$.

At finite $a$ and $m>0$ the lattice scheme becomes
mass-dependent too, as indicated by the function
$K_1(ma)=K_1(0)+\Delta K_1(ma)$ within $b_1$ above.
We denote the fermionic contribution to $v_{L,2}$ as
$n_fv_{L,2}^f$ and $V_{L,2}^f=v_{L,2}^f/(4\pi)^2$.
The latter, multiplied by $-R$, is displayed in Fig.~\ref{fig:massive}
for the example of Wilson fermions.
In the massless case we find the
expected logarithmic running, proportional to the fermionic contribution to
$\beta_0$ (straight line).
The offset at small $R$ is related to $K_1(ma)$, which we determine
by matching the lattice potential at $R\gg 1$ to the known
$\overline{MS}$ result.
$\Delta K_1$ is only unique in the limit
$m=0$ (where it vanishes). For massive quarks universality is lost
and $\Delta K_1$ will depend on the observable that is matched.
In the case of the position space potential (and force) we obtain the
asymptotic behaviour,
$\Delta K_1(ma)=0.0011(3)-K_1(0)-\ln(ma)/(3\pi)$ for $ma\rightarrow\infty$.
The logarithmic term is universal and guarantees
the massive fermions to decouple from the $\beta$ function.

We display the size of this correction, relative to $K_1(0)<0$ in
Fig.~\ref{fig:shift}. Note that in the limit $ma\rightarrow\infty$
the ratio $-\Delta K_1/K_1$ diverges towards
negative values. For the two $O(a)$ improved fermionic
actions the mass dependence is minimal while this effect can be
very significant for Wilson fermions and has to be
taken into account in any calculation
that uses the potential as an intermediate scheme.

As can be seen from Fig.~\ref{fig:massive}, at $r\ll m^{-1}$, i.e.\
$R\ll (ma)^{-1}$, the effective Coulomb coupling is screened with
the same logarithmic slope as in the massless case. At $r\gg m^{-1}$
the heavy quarks decouple and do not contribute to the running
anymore but just to the overall normalisation. It is here that
the un-quenched potential can be matched to the quenched one
by adjusting the coupling constant $\beta=3/(2\pi\alpha_L)$:
$\beta^{(n_f)}=\beta^{(0)}+\Delta\beta$,\
$\Delta\beta=n_f\left[\ln(Dma)+3\pi\Delta K_1(ma)\right]/(2\pi^2)+
O(\beta^{-1})$.
We find the numerical values,
$D=0.448(2), 0.0238(1)$ and $0.726(2)$, for Wilson, SW and
KS fermions, respectively. As
$ma\rightarrow\infty$ the $\Delta K_1$ term guarantees that
$\Delta\beta\rightarrow 0$ while $\Delta\beta$ diverges
as $ma\rightarrow 0$: in the light quark limit the quenched and un-quenched
theories decouple.

In
Fig.~\ref{fig:compbshift} we compare our perturbative prediction
to numerical data with $n_f=2$ Wilson~\cite{Bali:2000vr},
SW~\cite{Allton:2001sk} and KS~\cite{Tamhankar:2000ce} fermions,
obtained at lattice spacings $4<r_0/a<6.5$
and quark masses
$0.1\geq ma\geq 0.01$. The Wilson and SW results fall onto a universal
curve that differs by less than 10~\% from the prediction while
the KS results deviate much more and some $\beta$ dependence is
evident. Whether this is due to a slower convergence of the
perturbative series
or due to $n_f$ not being a multiple of four
is an open question.
The qualitative agreement between prediction and simulation 
for Wilson-type quarks indicates that,
at least at present masses, physics at hadronic scales is not
strongly affected by quark loops, which is consistent with the
phenomenological success of the quenched approximation.

\section{THE QCD RUNNING COUPLING}
\begin{figure}
\includegraphics[width=17pc]{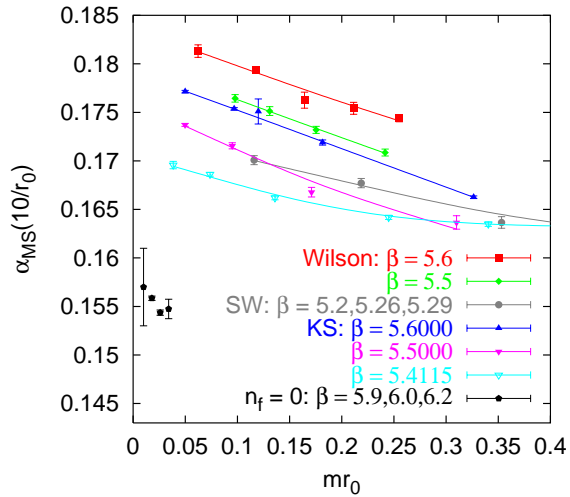}
\caption{QCD coupling for $n_f=2$.\vskip -.7cm}
\label{fig:alpha2}
\end{figure}

We determine the running coupling
$\alpha_{\overline{MS}}(3.401\,a^{-1})$
from the average plaquette, following the method
detailed in Ref.~\cite{Davies:1997mg}. The result is then converted into
units of $r_0$~\cite{Bali:2000vr,Allton:2001sk,Tamhankar:2000ce} and
numerically
evolved to the scale $\mu=10\,r_0^{-1}$ (Fig.~\ref{fig:alpha2})
via the four-loop $\beta$ function.
While the mass dependence of the plaquette is rather weak, the
term $\Delta K_1(ma)$ has a statistically significant impact
on the Wilson results at masses $ma> 0.02$.

The bottom-left $n_f=0$ pentagon
corresponds to the ALPHA collaboration result, 
$\alpha_{\overline{MS}}(\mu)=0.157(4)$, the other ones
have been
obtained from the average plaquette within a $\beta$-window that
covers the range of lattice spacings, spanned by the
dynamical simulations. We note that the $n_f=0$ values obtained by use
of the two different methods agree. All dynamical results
can be fitted to polynomials in
$mr_0$ (solid curves). The SW results~\cite{Allton:2001sk} have been
obtained at fixed $r_0$ while the two Wilson~\cite{Bali:2000vr}
and three KS~\cite{Tamhankar:2000ce} data sets
have been produced at fixed $\beta$.
We observe a significant lattice spacing dependence,
in contrast to the $n_f=0$ case. Extrapolating to the limits
$mr_0\approx 0.01$ and $a/r_0\rightarrow 0$, the latter linearly
for Wilson and quadratically for KS fermions, we obtain
the consistent estimate,
$\alpha^{(2)}_{\overline{MS}}(\mu)-
\alpha^{(0)}_{\overline{MS}}(\mu)=0.030(6)$.
Simulations at additional lattice spacings are mandatory
to control the
extrapolation to the continuum limit. Furthermore, numerical data
at $n_f\neq 0,2$ are essential, before
$\alpha^{(5)}_{\overline{MS}}(m_Z)$ can be predicted with
confidence. Once such results are available an error of $2~\%$
appears to be realistic.\\

\noindent {\bf ACKNOWLEDGMENTS}\\

G.B.\ is a Heisenberg Fellow (DFG grant Ba~1564/4-1).            
This work is supported by PPARC grants
PPA/G/O/1998/00559 and PPA/J/S/1998/00756
and the EU networks HPRN-CT-2000-00145 and
FMRX-CT97-0122.

\end{document}